\documentclass[aps, prl, twocolumn, showpacs, superscriptaddress, floatfix]{revtex4}
\usepackage{amsmath} 
\usepackage{amssymb}	
\usepackage{graphicx} 
\usepackage{times}
\usepackage{color}

\begin{document}
\newcommand\bbone{\ensuremath{\mathbbm{1}}}
\newcommand{\ul}{\underline}
\newcommand{\bp}{{\bf p}}
\newcommand{\vl}{v_{_L}}
\newcommand{\vc}{\mathbf}
\newcommand{\be}{\begin{equation}}
\newcommand{\ee}{\end{equation}}
\newcommand{\bk}{{{\bf{k}}}}
\newcommand{\bK}{{{\bf{K}}}}
\newcommand{\cE}{{{\cal E}}}
\newcommand{\bQ}{{{\bf{Q}}}}
\newcommand{\br}{{{\bf{r}}}}
\newcommand{\bg}{{{\bf{g}}}}
\newcommand{\bG}{{{\bf{G}}}}
\newcommand{\hbr}{{\hat{\bf{r}}}}
\newcommand{\bR}{{{\bf{R}}}}
\newcommand{\bq}{{\bf{q}}}
\newcommand{\hx}{{\hat{x}}}
\newcommand{\hy}{{\hat{y}}}
\newcommand{\hd}{{\hat{\delta}}}
\newcommand{\bea}{\begin{eqnarray}}
\newcommand{\eea}{\end{eqnarray}}
\newcommand{\ra}{\rangle}
\newcommand{\la}{\langle}
\renewcommand{\tt}{{\tilde{t}}}
\newcommand{\upa}{\uparrow}
\newcommand{\dna}{\downarrow}
\newcommand{\bS}{{\bf S}}
\newcommand{\vS}{\vec{S}}
\newcommand{\dg}{{\dagger}}
\newcommand{\pdg}{{\phantom\dagger}}
\newcommand{\tphi}{{\tilde\phi}}
\newcommand{\cf}{{\cal F}}
\newcommand{\ca}{{\cal A}}
\renewcommand{\ni}{\noindent}
\newcommand{\ct}{{\cal T}}
\newcommand{\brf}{\bar{F}}
\newcommand{\brg}{\bar{G}}
\newcommand{\jeff}{j_{\rm eff}}

\title{Highly-anisotropic exchange interactions of $j_{\rm eff}=1/2$ iridium moments on the fcc lattice in La$_2B$IrO$_6$ ($B$~$=$~Mg, Zn)}

\author{A. A. Aczel}
\altaffiliation{author to whom correspondences should be addressed: E-mail:[aczelaa@ornl.gov]}
\affiliation{Quantum Condensed Matter Division, Oak Ridge National Laboratory, Oak Ridge, TN 37831, USA}
\author{A. M. Cook}
\affiliation{Department of Physics, University of Toronto, Toronto, Ontario, Canada M5S 1A7}
\author{T. J. Williams}
\affiliation{Quantum Condensed Matter Division, Oak Ridge National Laboratory, Oak Ridge, TN 37831, USA}
\author{S. Calder}
\affiliation{Quantum Condensed Matter Division, Oak Ridge National Laboratory, Oak Ridge, TN 37831, USA}
\author{A.D. Christianson}
\affiliation{Quantum Condensed Matter Division, Oak Ridge National Laboratory, Oak Ridge, TN 37831, USA}
\author{G.-X. Cao}
\affiliation{Materials Science and Technology Division, Oak Ridge National Laboratory, Oak Ridge, TN 37831, USA}
\author{D. Mandrus}
\affiliation{Materials Science and Technology Division, Oak Ridge National Laboratory, Oak Ridge, TN 37831, USA}
\affiliation{Department of Materials Science and Engineering, University of Tennessee, Knoxville, TN 37996, USA}
\author{Y. B. Kim}
\affiliation{Department of Physics, University of Toronto, Toronto, Ontario, Canada M5S 1A7}
\affiliation{Canadian Institute for Advanced Research, Toronto, Ontario, M5G 1Z8, Canada}
\author{A. Paramekanti}
\altaffiliation{author to whom correspondences should be addressed: E-mail:[arunp@physics.utoronto.ca]}
\affiliation{Department of Physics, University of Toronto, Toronto, Ontario, Canada M5S 1A7}
\affiliation{Canadian Institute for Advanced Research, Toronto, Ontario, M5G 1Z8, Canada}
\date{\today}

\begin{abstract}
We have performed inelastic neutron scattering (INS) experiments to investigate the magnetic excitations in the weakly distorted face-centered-cubic (fcc) iridate double perovskites La$_2$ZnIrO$_6$ and La$_2$MgIrO$_6$, which are characterized by A-type antiferromagnetic ground states. The powder inelastic neutron scattering data on these geometrically frustrated $j_{\rm eff}=1/2$ Mott insulators provide clear evidence for gapped spin wave excitations with very weak dispersion.
The INS results and thermodynamic data on these materials can be reproduced by conventional Heisenberg-Ising models with significant uniaxial Ising anisotropy and 
sizeable second-neighbor ferromagnetic interactions. Such a uniaxial Ising exchange interaction is 
symmetry-forbidden on the ideal fcc lattice, so that it can only arise from the weak crystal distortions away from the ideal fcc limit.
This may suggest that even weak distortions in $j_{\rm eff}=1/2$ Mott insulators might lead to
 strong exchange anisotropies. More tantalizingly, however, we find an alternative viable explanation of the INS results in terms of
spin models with a dominant Kitaev interaction.
In contrast to the uniaxial Ising exchange, the highly-directional Kitaev interaction is a type of exchange anisotropy which is
symmetry-allowed even on the ideal fcc lattice.
The Kitaev model has a magnon gap induced by quantum order-by-disorder, while weak anisotropies of the Kitaev couplings 
generated by the symmetry-lowering due to lattice distortions, can pin the order and enhance the magnon gap.
Our findings highlight how even conventional magnetic orders in heavy transition metal oxides may 
be driven by highly-directional exchange interactions rooted in strong spin-orbit coupling.
\end{abstract}


\maketitle

\section{I. Introduction}
Magnetic materials with strongly-anisotropic exchange interactions often display exotic magnetic properties and allow fundamental theories of quantum magnetism to be tested in the laboratory \cite{81_yoshizawa, 83_nagler_2, 14_kinross}. Many interesting studies on these topics have concentrated on systems based on rare earths or 3$d$ transition metals with unquenched orbital angular momentum. For example, the classical spin ice pyrochlores Dy$_2$Ti$_2$O$_7$ and Ho$_2$Ti$_2$O$_7$ with net ferromagnetic nearest neighbor exchange and dipole interactions are well-described by an Ising Hamiltonian with the moments constrained to lie along the local $\la 111 \ra$ directions \cite{10_gardner}. Recent studies have shown that such classical spin-ice materials, with `two-in, two-out' magnetic ground states, support emergent monopole excitations \cite{08_castelnovo}, and there is ongoing work aimed at understanding 'quantum spin-ice' physics in materials such as Yb$_2$Ti$_2$O$_7$ \cite{11_ross}. 

Recently, a novel family of magnetic materials based on strong spin-orbit coupling (SOC) and the $d^5$ electron configuration, so-called $\jeff\!\!=\!\!1/2$ Mott insulators, have been attracting great interest \cite{08_kim}, as the relativistic entanglement of the orbital and spin degrees of freedom leads to unusual single ion wavefunctions \cite{09_jackeli}. Two different types of interactions for these wavefunctions have been considered in the ideal limit of a local cubic environment: superexchange mediated by a single anion via (a) a 90$^\circ$ bond (e.g. edge-sharing octahedra) and (b) a 180$^\circ$ bond (e.g. corner-sharing octahedra).

Superexchange through a 180$^\circ$ bond leads to a Hamiltonian with Heisenberg and anisotropic pseudodipolar terms \cite{09_jackeli}.
Resonant inelastic x-ray scattering (RIXS), performed to probe magnons in the single layer iridate Sr$_2$IrO$_4$, shows that its magnetic excitations are 
consistent with a dominant Heisenberg interaction \cite{12_kim}, with a small gap arising from spin-orbit-induced anisotropic couplings. 
However, RIXS on the bilayer iridate Sr$_3$Ir$_2$O$_7$ provides evidence for a large spin gap
ascribed to a significant interplane pseudodipolar term for spins within the bilayer, leading to the 
spin gap and bandwidth of the magnons 
having comparable values of 92 and 70~meV respectively \cite{12_kim_2}. Such a pseudodipolar (Ising-like) term is symmetry-allowed in a tetragonal crystal.
Its magnitude is thought to be large in Sr$_3$Ir$_2$O$_7$ due to the proximity to a metal-insulator transition and a sizable tetragonal distortion of the IrO$_6$ octahedra,
which increase the impact of the Hund's coupling in the
superexchange process as well as promote greater mixing of $j_{\rm eff}\!=\! 1/2$ and $j_{\rm eff}\!=\! 3/2$ states \cite{12_kim_2}.

Superexchange through a 90$^\circ$ bond is even more intriguing. In this geometry, there are two different exchange paths connecting magnetic ions. Projecting the $t_{2g}$ hopping onto the $j_{\rm eff}=1/2$ states leads to destructive interference of the hopping amplitudes, suppressing the usual superexchange. Incorporating Hund's coupling 
can then lead to a Hamiltonian where anisotropic exchange terms dominate \cite{09_jackeli}. For the special case of the two-dimensional (2D) honeycomb lattice, 
the resulting Hamiltonian is called the `Kitaev model' \cite{06_kitaev} and it is exactly solvable, with a quantum spin liquid ground state and emergent Majorana fermion excitations.

A true example of a Kitaev spin liquid has, however, remained elusive, as experimental realizations of the 2D honeycomb lattice such as $\alpha$-Li$_2$IrO$_3$ \cite{12_singh} Na$_2$IrO$_3$ \cite{10_singh}, and $\alpha$-RuCl$_3$ \cite{14_plumb} are characterized by magnetically {\it ordered} ground states \cite{12_ye, 15_sears} due to non-negligible Heisenberg, off-diagonal, or further neighbor exchange couplings \cite{10_chaloupka, 11_kimchi, 14_rau, 14a_perkins}. On the other hand, recent Raman scattering \cite{14b_perkins, 15_sandilands} and inelastic neutron scattering (INS) \cite{14_knolle, 15_banerjee} measurements have found evidence for strong Kitaev interactions in $\alpha$-RuCl$_3$, suggesting the ordered ground state may be proximate to a quantum phase transition into the Kitaev spin liquid ground state. Finally, recent experiments on the 3D honeycomb polymorphs $\beta/\gamma$-Li$_2$IrO$_3$ \cite{15_takayama, 14_modic} have uncovered complex spiral orders \cite{14_biffin, 14_biffin_2}, also ascribed to significant Kitaev exchange \cite{15_lee, 15_kimchi}.

While the quest to find an experimental example of a Kitaev spin liquid continues, a parallel effort is underway to characterize the magnetic properties of other $\jeff\!\!=\!\!1/2$ Mott insulators. Systems of particular interest have superexchange mediated by anions through 90$^\circ$ bonds, as they are prime candidates to host Kitaev-type exchange interactions \cite{14_kimchi}. Exploring cases in which the magnetic ion coordination number is different from the three-fold coordination of the honeycomb motif in previously studied
materials should yield further insights into the role of Kitaev interactions in quantum magnetism on other lattices.

Motivated by this background, we study $\jeff\!\!=\!\!1/2$ Mott insulating double perovskites (DPs) La$_2B$IrO$_6$ ($B=$Mg, Zn) \cite{95_currie, 13_cao, 15_zhu}, with Ir$^{4+}$ ions on the quasi-face-centered cubic (quasi-fcc) lattice. In these materials, the local octahedral environment of the Ir$^{4+}$ ions is very close to the cubic limit, and the larger Ir-Ir distance compared with $ABO_3$ perovskites leads to a Mott insulator, thus suggesting that the $\jeff\!=\!1/2$ description is appropriate. Although the DP structure does not feature direct edge-sharing IrO$_6$ octahedra, it has multiple interfering Ir-O-O-Ir paths.
As discussed in detail in the context of the triangular lattice iridate material Ba$_3$IrTi$_2$O$_9$, this
can suppress oxygen-mediated hopping of the
$j_{\rm eff}=1/2$ states and lead to significant Kitaev interactions \cite{15_trebst,15_catuneanu,15_jackeli}.
Since we can simply view the fcc crystal of Ir$^{4+}$ ions as a stacking of such triangular lattices along the $\{111\}$ direction, the
analysis in these works also applies to La$_2B$IrO$_6$.
Indeed, the fcc lattice has been theoretically proposed as a potential venue for hosting Kitaev interactions \cite{14_kimchi}.

The DP fcc structure has new features beyond previous, experimentally-studied, honeycomb-based materials: twelve-fold coordinated Ir sites, strong geometric frustration, and a larger Ir-Ir distance weakening direct Heisenberg exchange. This motivates us to explore the nature of magnetism in these materials in detail. The significance of the Kitaev interactions in La$_2B$IrO$_6$ is not at all evident from the observed magnetic ordering. Indeed, as explained in detail below, both materials exhibit A-type (Type-I) antiferromagnetic (AFM) ordering, with transition temperatures  $T_N$~$=$~12~K for La$_2$MgIrO$_6$ and 7.5~K for La$_2$ZnIrO$_6$ \cite{13_cao, 15_zhu, note}. 
Such commonly observed magnetic order on the fcc lattice can arise purely from a nearest-neighbor Heisenberg AFM exchange interaction.
However, we note that in this case the corresponding frustration parameter $f$, defined as $\Theta_{CW}/T_N$, the ratio of the Curie-Weiss temperature 
to the AFM ordering temperature, is  expected to be large, with $f \approx 9$ for spin-$1/2$ magnets, 
whereas the experimentally determined Curie-Weiss temperatures $\Theta_{CW} 
\approx -24$K for La$_2$MgIrO$_6$ and $\Theta_{CW} \approx -3$K for La$_2$ZnIrO$_6$ yield
small frustration values, $f \approx 2$ and 0.4 respectively \cite{13_cao}. One possible way to strongly suppress frustration is to
have large ferromagnetic second neighbor coupling \cite{98_seehra, 01_lefmann}. However,
as shown in recent theoretical work, even the symmetry-allowed nearest neighbor AFM Kitaev coupling on the 
fcc lattice cooperates with the nearest neighbor AFM Heisenberg exchange to greatly
stabilize A-type AFM. Interestingly, in the regime where this Kitaev interaction dominates, it leads to low $f$ values \cite{15_cook}. Given the strong SOC in these
materials,
this warrants a 
further exploration of the possible role played by 
such spin-orbit-induced directional exchange interactions on quantum magnetism in DPs.

\begin{figure*}[t]
\centering
\scalebox{0.46}{\includegraphics{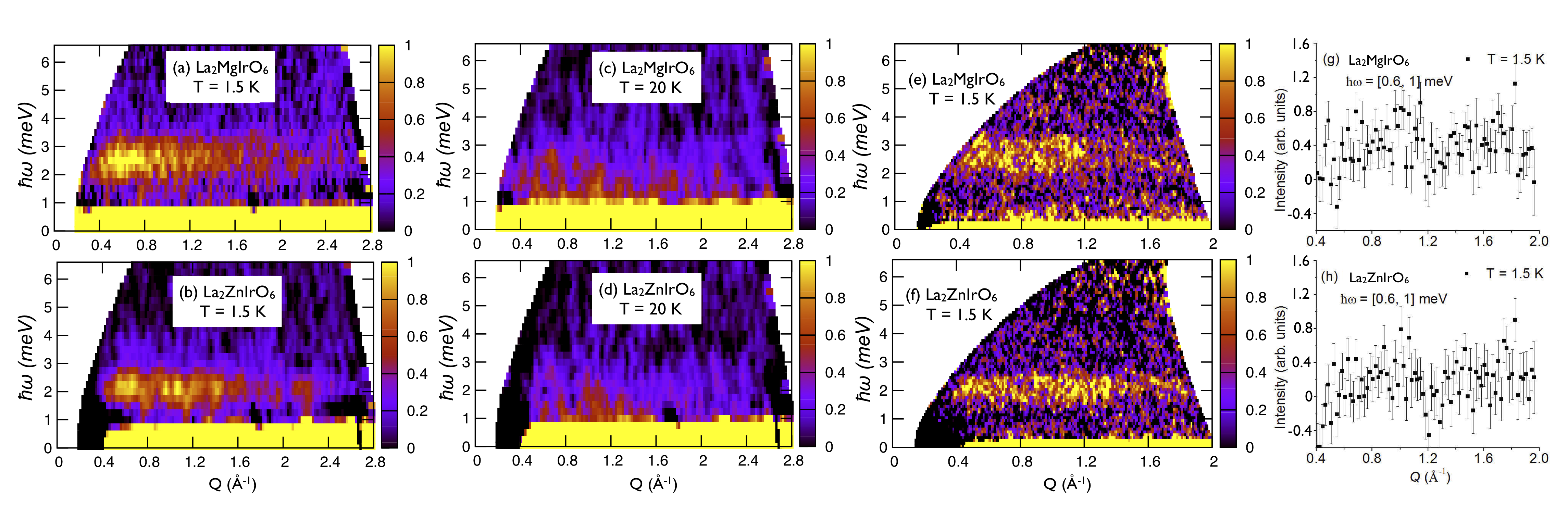}}
\caption{\label{HYSPEC} (Color online) Color contour plots of the coarse energy resolution HYSPEC data for (a) La$_2$MgIrO$_6$ and (b) La$_2$ZnIrO$_6$ with $E_i$~$=$~15~meV and $T$~$=$~1.5~K. Inelastic modes with weak dispersion are clearly observed for both materials. The color contour plots with $E_i$~$=$~15~meV and $T$~$=$~20~K shown in panels (c) and (d) indicate that these modes disappear above the respective $T_N$'s of 12~K and 7.5~K for the Mg and Zn systems, which suggests that they have a magnetic origin. Panels (e) and (f) depict similar color contour plots for La$_2$MgIrO$_6$ and La$_2$ZnIrO$_6$ at $T$~$=$~1.5~K, but with better energy resolution arising from the choice of $E_i$~$=$~7.5~meV. The same excitations are visible in these inelastic spectra. Note that the lowest-$Q$ regions in the color contour plots show no intensity; this issue results from a background oversubtraction of the direct beam, which is a consequence of the strong neutron absorption of Ir in the samples. Panels (g) and (h) show constant-$\hbar \omega$ cuts through the $T$~$=$~1.5~K fine-resolution datasets with an integration range of [0.6, 1] meV. The lack of increased intensity near the magnetic zone center $Q$~$=$~0.79~\AA$^{-1}$ for each system indicates that the magnetic excitations are fully-gapped.}
\end{figure*}

The smoking gun signature of any strong anisotropic couplings is most clearly encoded in the quantum spin fluctuations, and it reveals itself in the magnon spectrum. In this paper, we present results from an INS study of the magnetic excitations in La$_2B$IrO$_6$. Typically, INS is the most powerful technique to probe magnetic excitations in crystals. However, INS generally has severe limitations in most iridates due to an unfavorable magnetic form factor and the strong neutron absorption cross-section of the Ir nuclei, rendering RIXS as the tool of choice to study magnons \cite{12_kim, 12_kim_2, 13_gretarsson}. Remarkably, as shown in Fig.~\ref{HYSPEC}, we find that La$_2B$IrO$_6$ exhibit a clearly observable INS signal, revealing {\it gapped, highly non-dispersive} magnons. It is likely that the non-dispersive nature of the magnons is what leads to a higher intensity over a small
energy window, rendering them clearly visible in the INS measurements unlike for most other iridates.
Our INS work is important since RIXS does not yet possess the meV resolution to study low energy magnons in a strong Mott insulator. 

A comparison of our INS results with theoretical calculations shows that we can describe the data using conventional Heisenberg-Ising models with a uniaxial
Ising exchange anisotropy. However, the uniaxial Ising exchange interaction is symmetry-forbidden on the ideal
fcc lattice. This suggests that for such a Heisenberg-Ising model to provide a viable explanation of the data, even weak monoclinic distortions in these
 $j_{\rm eff}=1/2$ Mott insulators must, remarkably, be capable of inducing strong exchange anisotropies. 
We also discuss alternative spin models with dominant Kitaev interactions which are shown to capture the dynamical spin correlations in these
materials. In contrast to the uniaxial Ising term, the Kitaev interaction is symmetry-allowed even on the ideal fcc lattice, and 
weak lattice distortions may induce small anisotropies of the Kitaev coupling.
We thus propose that the Kitaev exchange might provide a more natural explanation for the combined
INS and thermodynamic observations in these materials. Our study 
suggests that even the {\it conventional} A-type AFM order in these geometrically-frustrated magnets may ultimately be selected by highly-anisotropic exchange interactions resulting from the strong spin-orbit coupling of heavy transition metal ions.

This paper is organized as follows: We begin with a discussion of the crystal structure and magnetic ordering patterns for La$_2B$IrO$_6$ in Sections II and III. We then present the INS data on these materials in Section IV which show evidence for gapped spin waves. In Section V,
we discuss the most likely phenomenological spin wave models needed to describe these results. Section VI discusses the relationship of these models to the experimental
data, providing estimates of exchange couplings which can semi-quantitatively reproduce the INS results.
Section VII discusses experimental estimates of $B/B'$ site-mixing disorder and its possible qualitative impact on the modelling.
Finally, Section VIII concludes with a summary of our work and implications for other materials.

\section{II. Crystal structure}

\begin{figure}[ht]
\centering
\scalebox{0.42}{\includegraphics{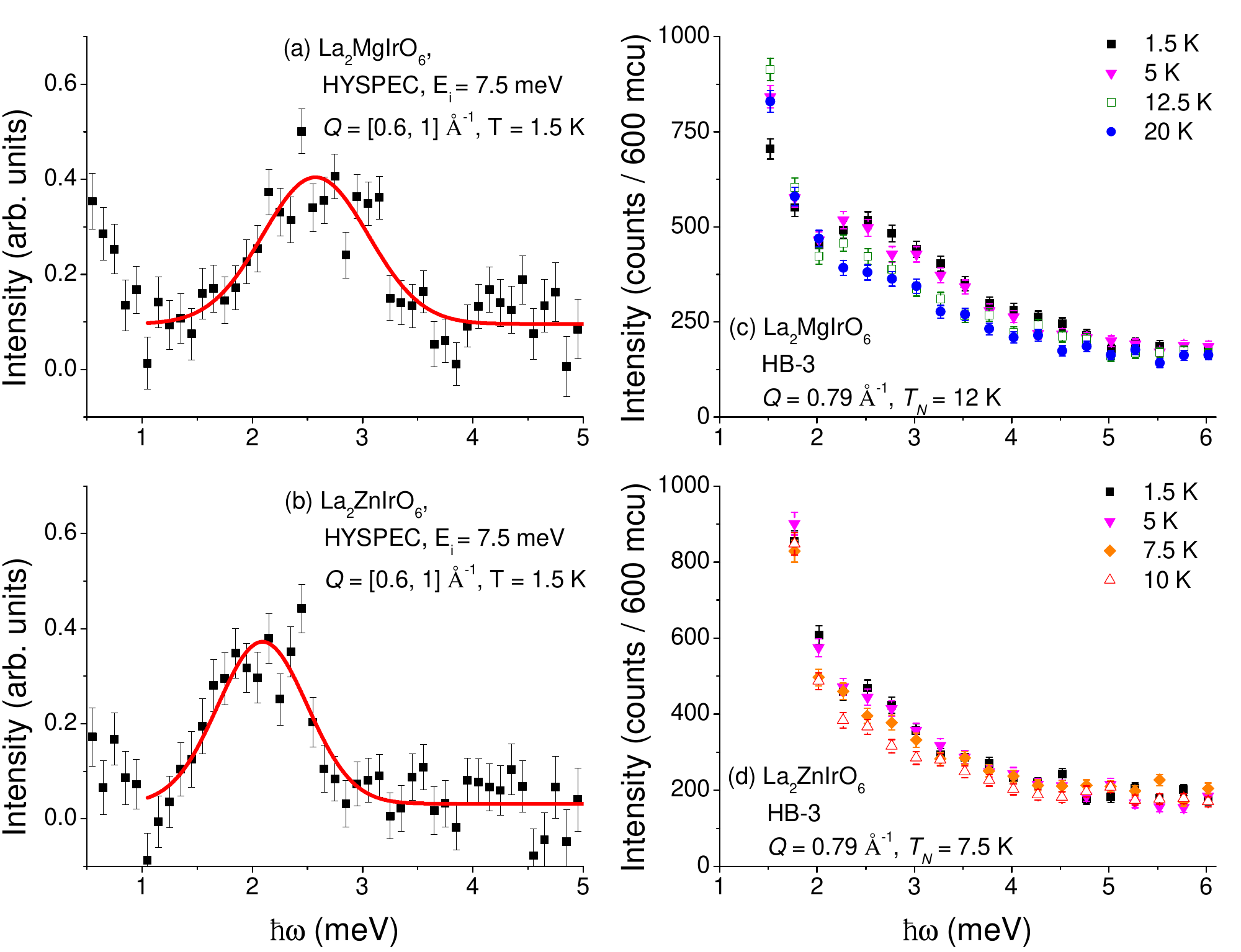}}
\caption{\label{HB3} (Color online) (a), (b) Constant-$Q$ cuts through the HYSPEC data shown in Fig.~\ref{HYSPEC}(a) and (b), integrated over $Q\!=\! [0.6,1]$\AA$^{-1}$. These cuts clearly show that the inelastic modes are gapped in each case. Gaussian fits are superimposed on the data, which were used to determine the peak positions. For La$_2$MgIrO$_6$, the mode is centered at $\hbar \omega = 2.57(4)$~meV, and for La$_2$ZnIrO$_6$, it is centered at $2.09(3)$~meV. (c), (d) HB-3 constant-$Q$ scans with $Q$~$=$~0.79~\AA$^{-1}$~for La$_2$MgIrO$_6$ and La$_2$ZnIrO$_6$ at selected temperatures. The spin gaps close around $T_N$ for each compound, indicating that the modes have a spin wave origin. Note that 1 mcu (monitor count unit) $\approx$ 10000 and 11000 monitor counts for the Mg and Zn data respectively.}
\end{figure}

Ordered double perovskites with the general formula $A_2BB'$O$_6$ ideally crystallize in a cubic structure, with the $B$ and $B'$ ions occupying two interpenetrating fcc sublattices. La$_2$MgIrO$_6$ and La$_2$ZnIrO$_6$ crystallize in the lower symmetry, monoclinic space group {\it P$2_1$/n}, arising from small structural distortions to the cubic structure. The unit cell associated with the {\it P$2_1$/n} space group is a superstructure of the primitive cubic unit cell, which can be approximately indexed in tetragonal notation due to the extremely weak monoclinic distortions. Assuming that $\hat{x}$, $\hat{y}$, and $\hat{z}$ are aligned with the three fcc crystallographic directions, the relationships between the tetragonal and fcc lattice constants are as follows: $\vec{a}_{\rm t}$~$=$~$a_{\rm fcc}(\hat{x}\pm\hat{y})/2$ and $\vec{c}_{\rm t}$~$=$~$a_{\rm fcc} \hat{z}$. For La$_2B$IrO$_6$, previous x-ray diffraction studies \cite{13_cao} have shown that $a_{\rm fcc}$~$\approx$~7.9~\AA.

We now make an effort here to quantify the magnitudes of the monoclinic structural distortions for the IrO$_6$ octahedra of La$_2B$IrO$_6$. The distortions have two main effects: the rotation of the octahedra about both the cubic [110] and c-axes, and the deformation of the Ir$^{4+}$ local environment away from ideal cubic. The rotation angles of the IrO$_6$ octahedra can be determined according to Ref.~\cite{86_groen} by using the refined atomic fractional coordinates and the Glazer notation discussed in Refs.~\cite{72_glazer, 97_woodward}. We find that the IrO$_6$ octahedra have global rotations of 13$^\circ$ and 14$^\circ$ for the Mg and Zn systems respectively about the cubic [110] axis, and rotations of 9$^\circ$ and 11$^\circ$ respectively about the c-axis that are staggered between adjacent ab-layers. The deformation of the IrO$_6$ octahedra can be quantified by considering the different Ir-O bond lengths and O-Ir-O bond angles. From the structural refinements reported in Ref.~\cite{13_cao}, we find that the six Ir-O bond lengths are within 1\%~of each other for both materials, and all O-Ir-O bond angles are within 3.5$^\circ$ (1.5$^\circ$) of 90$^\circ$ and 180$^\circ$ for the Mg (Zn) system. This implies a nearly-ideal local cubic environment for the Ir$^{4+}$ ions, which is consistent with $\jeff\!\!=\!\!1/2$ single ion ground states for La$_2B$IrO$_6$. 

\section{III. Magnetic ordering}

For La$_2$MgIrO$_6$, magnetization measurements show no evidence for a net ferromagnetic (FM) moment, while neutron powder diffraction work \cite{13_cao} finds a magnetic Bragg peak at $Q$~$=$~0.79~\AA$^{-1}$  corresponding to A-type AFM order. These combined results are consistent with a magnetic propagation vector of $\vec{k}$~$=$~(0.5 0.5 0)$_{\rm t}$, indicative of FM planes stacked along the [110]$_{\rm t}$ direction. Although the data do not determine the moment direction unambiguously, electronic structure calculations \cite{13_cao} predict that the moments lie predominantly in the FM planes (A-II type AFM in the notation of Ref.~\cite{15_cook}). 

For La$_2$ZnIrO$_6$, magnetization measurements find evidence for a net FM moment, while neutron diffraction again detects \cite{13_cao} a magnetic Bragg peak at $Q$~$=$~0.79~\AA$^{-1}$. These findings are consistent with a canted A-type AFM characterized by a $\vec{k}$~$=$~0 propagation vector, which defines the c-axis as the FM plane stacking direction. The magnetic Bragg peak is then uniquely indexed as (001)$_{\rm t}$. The observation of this peak, combined with the absence of the (100)$_{\rm t}$ and (010)$_{\rm t}$ peaks, strongly implies that the ordered moments lie predominantly in the FM planes. Thus, the A-type AFM in this system also corresponds to A-II. The spin canting in La$_2$ZnIrO$_6$ arises from the small, staggered IrO$_6$ octahedral rotations ($\sim$~11$^\circ$).

\section{IV. Inelastic neutron scattering results} 

We next present our new results on the magnetic excitations associated with the ordered phases of these materials. Inelastic neutron scattering data were collected on previously synthesized \cite{13_cao} powder samples of La$_2B$IrO$_6$ at the HYSPEC spectrometer of the Spallation Neutron Source, Oak Ridge National Lab (ORNL). The powder samples were loaded in Al annular cans to minimize neutron absorption. All data were collected using incident energies of $E_i$~$=$~7.5 and 15 meV, with corresponding Fermi chopper frequencies of 240 and 300~Hz, resulting in instrumental energy resolutions of 0.3 and 0.7 meV (Gaussian full-width half-maximum [FWHM]) respectively at the elastic line. A He cryostat was used to achieve a base temperature of 1.5~K. Empty Al annular can measurements were subtracted from all the HYSPEC data presented in this work, so the Al scattering contribution to the sample spectra would be minimized. INS data for these systems were also collected using a He cryostat on the thermal triple axis spectrometer HB-3 at the High Flux Isotope Reactor of ORNL. A collimation of 48'-60'-60'-120' and a fixed final energy of $E_f$~$=$~14.7~meV were used to achieve an energy resolution of 1.2~meV at the elastic line (Gaussian FWHM).

Fig.~\ref{HYSPEC}(a) and (b) depict color contour plots of the coarser energy resolution $E_i$~$=$~15~meV HYSPEC data at $T$~$=$~1.5~K, where a nearly-dispersionless excitation band is visible for both materials. Note that the lowest-$Q$ regions in these plots show no intensity; this issue results from a background oversubtraction of the direct beam, which is a consequence of the strong neutron absorption of Ir in the samples. The observed excitations are clearly magnetic in origin, as they decrease in intensity with increasing $Q$. They are also no longer visible above $T_N$, with only quasi-elastic scattering remaining, as shown in panels (c) and (d). Color contour plots of the finer energy resolution $E_i$~$=$~7.5~meV HYSPEC data are presented in Fig.~\ref{HYSPEC}(e) and (f). The same excitations are still present in the data, which indicates that they are not spurious in nature. Fig.~\ref{HYSPEC}(g) and (h) show constant-$\hbar \omega$ cuts through the fine-resolution data for the two materials, with an integration range of [0.6, 1] meV. We find no evidence for enhanced intensity near the $Q$~$=$~0.79~\AA$^{-1}$~magnetic Bragg peaks, which suggests that these excitations are fully-gapped. Finally, as shown in Fig.~\ref{HB3}(a) and (b), a low-$T$ constant-$Q$ 
cut through the fine-resolution data centered about the magnetic zone center $Q$~$=$~0.79~\AA$^{-1}$~reveals that the central position of the mode is $2.57(4)$~meV for La$_2$MgIrO$_6$ and $2.09(3)$~meV for La$_2$ZnIrO$_6$, with a FWHM in each case of $\sim 1$ meV.

We have also carried out detailed temperature-dependent measurements on the thermal triple axis spectrometer HB-3. Fig.~\ref{HB3}(c) and (d) present constant-$Q$ scans at a magnetic zone center ($Q$~$=$~0.79~\AA$^{-1}$) for La$_2$MgIrO$_6$ and La$_2$ZnIrO$_6$ respectively. The two panels provide strong evidence that the spin gaps close around $T_N$ in each case, with spectral weight shifting down to lower energies with increasing $T$. This observed temperature-dependence of the modes indicates that they likely correspond to spin waves. In fact, a crystal field interpretation can be ruled out by considering the typical single ion energy level scheme for $j_{\rm eff}\!=\! 1/2$ Mott insulators. For this class of materials, the lowest-lying excited state to the $j_{\rm eff}\!=\! 3/2$ band is separated from the $j_{\rm eff}\!=\! 1/2$ ground state by 3$\lambda_{SO}$/2, where $\lambda_{SO}$ is the spin-orbit coupling constant. The typical energy scale for this crystal field excitation is on the order of 100's of meV due to large $\lambda_{SO}$ \cite{12_liu, 13_gretarsson_2, 15_banerjee}, which is certainly incompatible with the energy scale of the magnetic modes observed here. With the magnetic excitations for La$_2B$IrO$_6$ now unambiguously identified as spin waves, we turn to a theoretical modeling of these results.  

\section{V. Theory}

The observed spin gap and weakly-dispersive spectra in the INS measurements, with the gap being comparable to or even larger than the magnon bandwidth, are 
suggestive of a nearly Ising-like exchange. Below, we examine the possible origins of this large exchange anisotropy in 
La$_2B$IrO$_6$.

The strategy we follow to construct the spin model
is very similar to that for other perovskite iridates \cite{09_jackeli}: namely, we start by assuming an ideal lattice (fcc), with no rotations or tilts of the IrO$_6$
octahedra, and consider all possible symmetry-allowed exchange couplings.
In this ideal fcc limit, the uniaxial Ising exchange interaction, which is the simplest anisotropic
exchange interaction typically used to model such gapped
magnon spectra, is {\it symmetry-forbidden}. However, even in this ideal limit, the highly-directional Kitaev interaction and an off-diagonal symmetric exchange
are allowed by symmetry \cite{15_cook}.

Moving away from the ideal fcc limit, the weak monoclinic distortions, arising from the rotations and tilts of the IrO$_6$ octahedra,
permit a large variety of new exchange terms just on symmetry grounds. 
A set of such exchange couplings induced by non-cubic distortions have been discussed in previous work in
the context of double perovskites \cite{11_dodds,14_ishizuka}. However, given the large number of new terms and the limited data,
we choose to focus on the simplest case of a tetragonal crystal distortion \cite{14_ishizuka} 
which allows for only a subset of all the terms; in particular, we focus on two new anisotropies which are symmetry-allowed under such a distortion. First, the tetragonal distortion allows for the Kitaev interaction to be different in one crystal plane compared with the other two orthogonal planes. Second, the tetragonal distortion picks a unique axis, thus now
permitting a uniaxial Ising interaction.

Below, we study the spin wave dispersion in such spin models and show that they can provide a reasonable description of the INS data. For ease of
presentation, we begin by first
discussing the conventional and simpler Heisenberg-Ising models with uniaxial Ising exchange, which might apply to non-ideal fcc lattices, 
followed by a study of the Kitaev-dominant models which begin from the ideal fcc limit. We then argue
that the Kitaev-dominant models appear to provide a more reasonable explanation of the combined INS data 
and previous measurements of $\Theta_{CW}$ and $T_N$.

\subsection{A. Heisenberg-Ising models with uniaxial Ising exchange}

\begin{figure*}
\centering
\scalebox{0.65}{\includegraphics{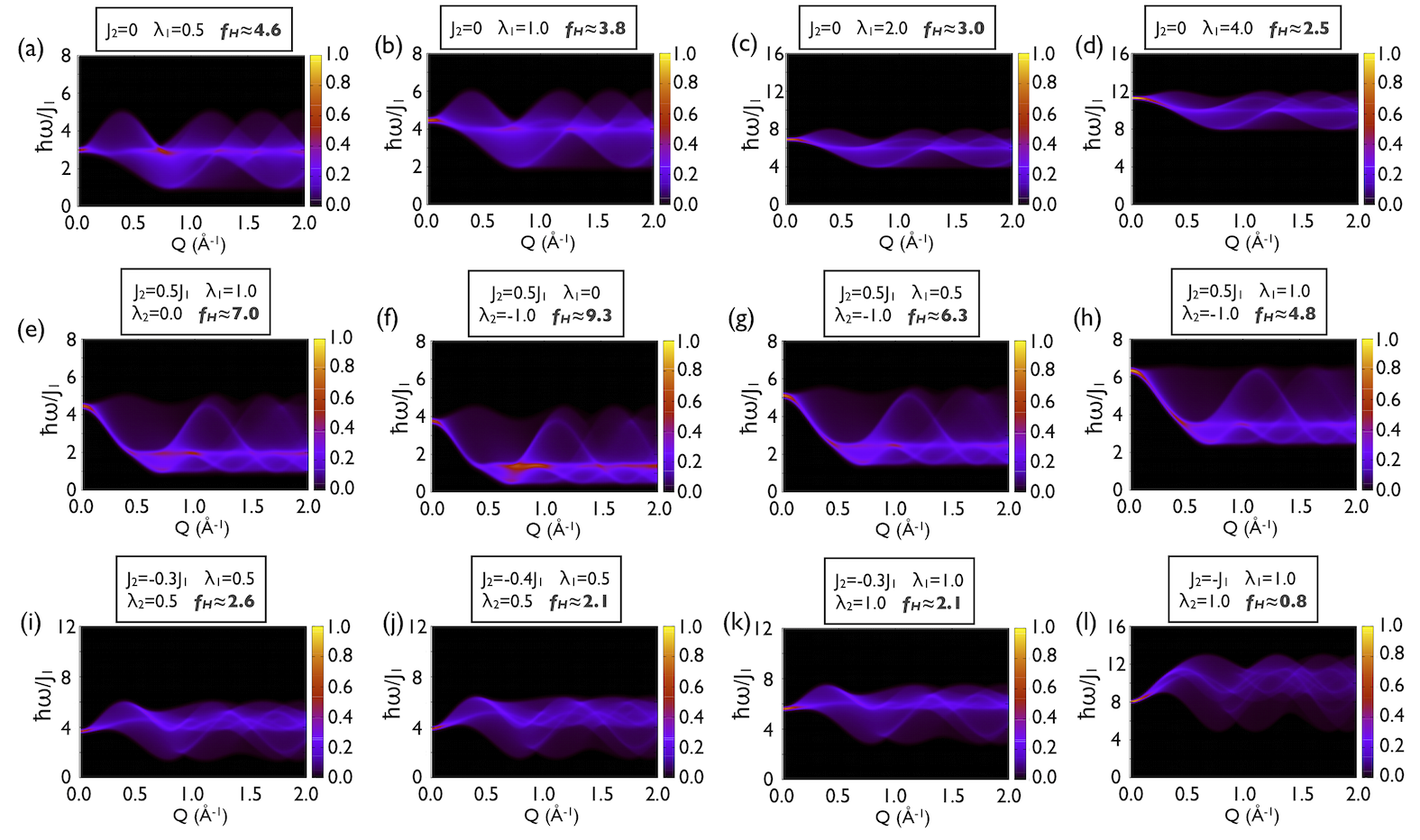}}
\caption{\label{theoryheis} (Color online) Theoretical powder-averaged $S(Q,\omega)$ for the Heisenberg-Ising models around the A-II AFM ground state considered in the text within linear spin-wave theory with
nearest neighbor AFM Heisenberg $J_1=1$, various second neighbor Heisenberg exchange $J_2$, and varying degrees of uniaxial anisotropies $\lambda_{1,2}$ in the 
first and second neighbor exchanges. 
We set $J_{1,2}^y\!=\!J^z_{1,2}\!=\!J_{1,2}$ and $J_{1,2}^x\!=\!(1\!+\! \lambda_{1,2}) J_{1,2}$. (a-d) Models with only first neighbor
exchange, with $J_2=0$. The corresponding frustration parameters $f_H$ are also shown, and range from $f_H \approx 9$ for the isotropic case with $\lambda_1=0$
to $f_H \approx 2$ for extreme anisotropy $\lambda_1 \gg 1$. (e-h) Heisenberg-Ising models with AFM second neighbor exchange, $J_2 > 0$, and varying degrees of anisotropy. 
(i-l) Heisenberg-Ising models with FM second neighbor exchange, $J_2 < 0$, and varying degrees of anisotropy. 
We have incorporated a small energy broadening to account for the finite instrumental energy resolution.}
\end{figure*}

Let us begin with the conventional and phenomenological
viewpoint that the spin gaps in many of these DPs simply arise from uniaxial Ising interactions. We emphasize that such uniaxial Ising couplings are {\it symmetry-forbidden} on the 
ideal fcc lattice; however, they may be permitted in the presence of
distortions away from the ideal fcc limit. Limiting ourselves to short-range exchange, such a Heisenberg-Ising Hamiltonian, where different 
spin components interact with different strengths, is given by
\bea
\!\! H_{HI} =   
\!\!\! \sum_{\la \br\br'\ra} J^\mu_1 \!S^\mu_\br S^\mu_{\br'}
+ 
\!\!\!  \sum_{\la\la \br\br'\ra\ra} J^\mu_2 \!S^\mu_\br S^\mu_{\br'}
\eea
where $\la.\ra$ and $\la\la. \ra\ra$ refer to nearest and next-nearest neighbor sites. We set $J_{1,2}^y\!=\!J^z_{1,2}\!=\!J_{1,2}$ and 
$J_{1,2}^x\!=\!(1\!+\! \lambda_{1,2}) J_{1,2}$.
For $J_2 > 0$, we select $\lambda_1\!>\! 0$ and
$\lambda_2\!<\! 0$ which have been shown \cite{15_taylor_2} to stabilize A-II type AFM when the anisotropy is strong enough to overcome the AFM $J_2$.
For $J_2 < 0$, we fix $\lambda_{1,2} \!>\! 0$.

The powder-averaged dynamic structure factors $S(Q,\omega)$ for these models are shown in Fig.~\ref{theoryheis} for various values of $J_2/J_1$ and anisotropies
$\lambda_1,\lambda_2$.
It is clear from Fig.~\ref{theoryheis} that $S(Q,\omega)$ exhibits a high intensity band of gapped dispersionless excitations which resembles the INS data, for 
either sign of $J_2$ and varying degrees of anisotropy. 
We also show the corresponding frustration parameters $f_H$. We obtain them from the powder averaged 
$\theta_{CW}= - J_1(3+\lambda_1)- J_2 (3+\lambda_2)/2$,
with $T_N$ determined from classical Monte Carlo simulations rescaled by $S(S+1)$ to account for quantum effects.
We note that for models with AFM $J_2\! \geq \! 0$, the frustration parameter $f_H$ is large even for significant anisotropy $\lambda_{1,2} \! \sim\! {\cal O}(1)$,
approaching the experimentally relevant regime of small frustration $f_H \! \sim \! 2$ only for extremely strong uniaxial anisotropy.
However, is it reasonable to assume such an extreme uniaxial Ising anisotropy $\lambda_{1,2} > {\cal O}(1)$?

To answer this question, we begin by noting that models with large uniaxial Ising exchange anisotropy adequately describe qualitatively-similar INS spectra in certain
$3d$-transition metal ion based magnets such as CsCoCl$_3$ and CsCoBr$_3$ \cite{83_nagler, 78_tellenbach}.
In these two Co$^{2+}$ systems, the 
comparable values for SOC ($\sim\!28$~meV for CsCoCl$_3$ \cite{95_nguyen}) and the non-cubic crystal field distortion parameter $\delta$ in the single ion Hamiltonian lead to the large uniaxial Ising anisotropies ($\lambda \approx 7-8$ \cite{83_nagler}). In La$_2B$IrO$_6$, a similarly large uniaxial Ising anisotropy with a single ion origin should 
be accompanied by a complete 
breakdown of the $\jeff\!\!=\!\!1/2$ description, with a significant modification of the isotropic single ion ground state wavefunctions \cite{12_liu}. However, this
breakdown is unlikely, as
the SOC scale is expected to be on the order of 100's of meV
and much larger than $\delta$. Indeed, the Ir$^{4+}$ local octahedral environments in La$_2B$IrO$_6$ are amongst the closest to ideal cubic of any $\jeff\!\!=\!\!1/2$ Mott insulating candidates \cite{12_liu}. Furthermore, recent RIXS measurements have validated a $\jeff\!\!=\!\!1/2$ ground state for Na$_2$IrO$_3$ \cite{13_gretarsson_2}, which exhibits much larger IrO$_6$ octahedral distortions.

Among other iridates, RIXS work on the stacked bilayer material Sr$_3$Ir$_2$O$_7$, 
has shown that the spin waves exhibit a significant magnon gap \cite{12_kim_2}. This has been understood as a result of a significant uniaxial Ising
exchange ($\lambda$~$\sim$~1.4) between spins on adjacent planes in the bilayer which is comparable to the Heisenberg exchange.
The large magnitude of this term in Sr$_3$Ir$_2$O$_7$
arises from the sizable tetragonal distortions of the IrO$_6$ octahedra, and 
the proximity to a metal-insulator transition; this leads to increased mixing of $j_{\rm eff}=1/2$ and $j_{\rm eff}=3/2$ states
and a greater impact of Hund's coupling on intermediate states 
in the superexchange process \cite{12_kim}. Neither of these effects,
the strong tetragonal distortions or proximity to a metal-insulator transition, are applicable for La$_2B$IrO$_6$ since they are
in the Mott insulating regime \cite{13_cao} and have nearly cubic IrO$_6$ octahedra.

A more complete theory \cite{03_kuzmin,14_ishizuka} for the magnetic excitations of La$_2B$IrO$_6$ 
must incorporate antisymmetric Dzyaloshinskii-Moriya (DM) interactions arising from octahedral
rotations. However, previous work on Sr$_2$IrO$_4$ shows that the effect of octahedral rotations in these $j_{\rm eff}\!=\!1/2$ Mott insulators
may be largely accounted for by making
local spin rotations on the Heisenberg model \cite{09_jackeli}, so the Ir moments track the local octahedral rotations \cite{13_boseggia}.
This `rotated' Hamiltonian features DM interaction and additional terms, which
however does not change the spectrum since it is related by a unitary transformation to the Heisenberg model. 
This hidden Heisenberg symmetry \cite{09_jackeli, 11_senthil}
explains the existence of highly dispersive and nearly-gapless magnetic excitations in Sr$_2$IrO$_4$ 
\cite{12_kim, 12_fujiyama, 14_bahr}. In bilayer Sr$_3$Ir$_2$O$_7$, which has
staggered IrO$_6$ octahedral rotations about the c-axis of $\sim$~12$^\circ$ \cite{12_boseggia} and a large magnon gap,
it is important to note that 
the pseudodipolar interaction which is thought to be responsible for the gap would be symmetry-allowed even in the {\it absence} of octahedral rotations,
and therefore this term does not arise solely from these rotations \cite{12_kim}.
Thus, although La$_2$BIrO$_6$ with their {\it P2$_1$/n} structure have comparable IrO$_6$ octahedral rotations  to Sr$_3$Ir$_2$O$_7$ (albeit about 
two different crystallographic axes), these rotations alone cannot be responsible for any large new exchange couplings which
do not arise from already symmetry-allowed exchange couplings in the ideal fcc lattice.

The above arguments suggest that the regime of extreme and dominant uniaxial Ising anisotropy, $\lambda_{1,2} > {\cal O}(1)$, is 
unlikely for La$_2B$IrO$_6$, so this subset of models is no longer considered here. We also note that the Heisenberg-Ising models with $J_2 \!>\! 0$ and $\lambda_{1,2} \sim {\cal O}(1)$ yield frustration parameters, $f_{H} \!\sim\! 5$-$9$ (see Fig.~\ref{theoryheis}), significantly larger than 
the measured values $f \!\sim\! 0.4$-$2$ \cite{13_cao}, and therefore these models are unlikely to explain the data on La$_2B$IrO$_6$ either.
On the other hand, we find that choosing FM $J_2 < 0$ does lead to models with smaller frustration parameters.
In this case, as shown in Fig.~\ref{theoryheis}, we find that we can explain the magnon gap as well as the frustration parameter, but only by choosing
significant anisotropies $\lambda_{1,2} \sim {\cal O}(1)$.

Our main observation here is that within this
class of phenomenological Heisenberg-Ising models, we can only explain the combined INS and thermodynamic data by invoking significant uniaxial Ising anisotropy, 
as well
as significant FM second neighbor exchange. We next explore competing models with dominant Kitaev interactions which provide 
an unconventional exchange anisotropy, which is symmetry-allowed even in the ideal fcc limit.

\subsection{B. Models with dominant Kitaev exchange}
We have previously studied the classical phase diagram for ideal $\jeff\!=\!1/2$ fcc magnets \cite{15_cook}, keeping all symmetry-allowed nearest neighbor (NN) interactions including Heisenberg, Kitaev, and off-diagonal symmetric exchange \cite{86_halg,14_rau,15_cook}. A key finding, relevant to La$_2B$IrO$_6$, was that while the simple NN AFM Heisenberg model exhibits A-type AFM order, this exact same order is also favored in the regime of dominant AFM Kitaev exchange. 
By contrast, while a small  off-diagonal symmetric exchange 
does not significantly affect our results, a dominant value of this coupling leads to ordered moments pointing along the $\la 111 \ra$ direction 
\cite{15_cook} which does not agree with previous neutron diffraction work for La$_2B$IrO$_6$ \cite{13_cao}. 
One might distinguish between the regimes of weak and dominant Kitaev interactions by using the frustration parameter, $f$. For the NN Heisenberg AFM model on the fcc lattice, we estimated $f \approx 9$ for spin-$1/2$ moments \cite{15_cook}. However, these iridates exhibit robust AFM order, with experimental values for $f \lesssim 2$, suggesting that SOC-induced Kitaev interactions are large, suppressing frustration and enhancing 
$T_N$. This led us to propose a Heisenberg-Kitaev model on the ideal fcc lattice,
\bea
\!\!\!\!\!\! H \!\!&=&\!\! J_1 \sum_{\la \br \br' \ra} \vec S_\br \cdot \vec S_{\br'}+   J_2 \sum_{\la\la \br \br' \ra\ra} \vec S_\br \cdot \vec S_{\br'} + H_K\\
\!\!\!\!\!\! H_K \!\! &=& \!\! J_K \!\!\!  \sum_{\la \br\br'\ra_{yz}} \!\!\!\!  S^x_\br S^x_{\br'}+
  J_K \!\!\!  \sum_{\la \br\br'\ra_{xz}} \!\!\!\!  S^y_\br S^y_{\br'} +
     J_K \!\!\! \sum_{\la \br\br'\ra_{xy}}\!\!\!\! S^z_\br S^z_{\br'}
\eea
with $J_K \! \gg \! J_1, J_2$,
as a better starting point to describe the magnetism in La$_2B$IrO$_6$. Here, $\la \br \br' \ra$ and $\la\la \br \br' \ra\ra$ denote first and second neighbors on the
fcc lattice, while $\la \br \br' \ra_{yz}$ denotes nearest neighbors in the $yz$-plane (similarly for $xz,xy$). This model has a
powder averaged $\Theta_{CW} \!=\! -J_K - 3 J_1 - 3 J_2/2$, so if the Kitaev coupling is dominant then $J_K > 0$ 
is consistent with the reported $\Theta_{CW}\!<\!0$ for La$_2B$IrO$_6$ \cite{13_cao}. A classical Monte Carlo study \cite{15_cook} of this model for $J_2=0$ 
and $J_K /J_1 \gg 1$ showed that $f \! \approx \! 2$, in reasonable agreement with the data on La$_2$MgIrO$_6$. Here, we focus on the 
effects of quantum spin fluctuations and the resulting dynamic structure factor of such Kitaev-dominant models.

\begin{figure*}
\centering
\scalebox{0.45}{\includegraphics{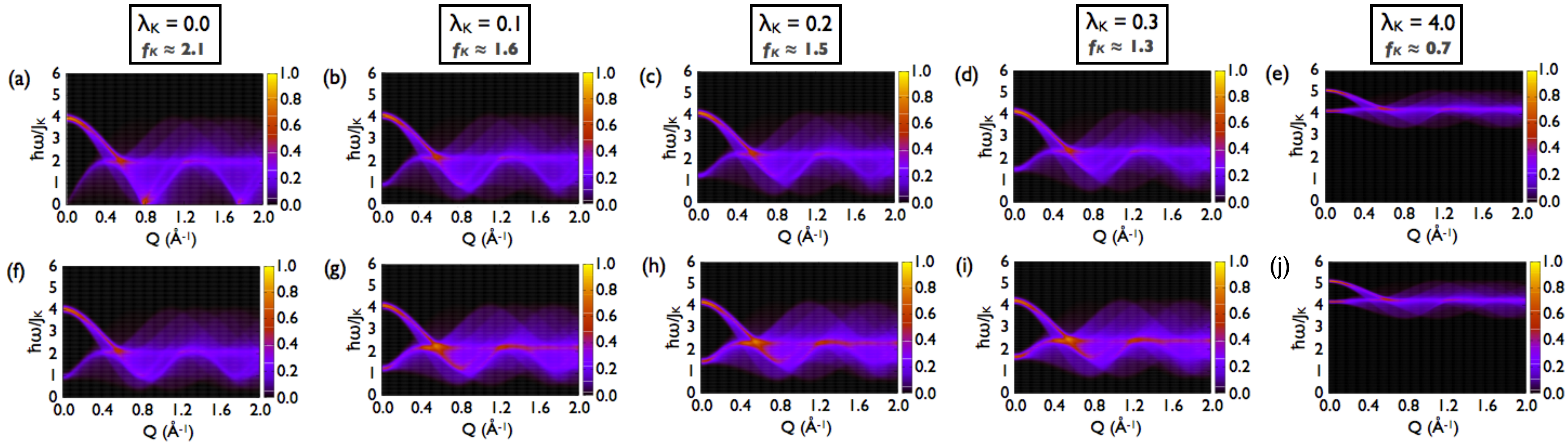}}
\caption{\label{theory} (Color online) Theoretical powder-averaged $S(Q,\omega)$ for the Kitaev models around the A-II AFM state for indicated
spatial anisotropy $\lambda_K$. (a-e) Kitaev model within linear spin wave theory for the ideal fcc limit ($\lambda_K\!=\! 0$, no spin gap) and weak
anisotropy ($\lambda_K\!=\! 0.4$). (f-j) Kitaev models incorporating magnon interactions which induce an order-by-disorder spin 
gap even for $\lambda_K\!=\!0$, and a slight enhancement of the 
spin gap for $\lambda_K \!>\! 0$. The corresponding frustration parameters $f_K$ are also shown, they range from $f_K \approx 2$ for the ideal fcc lattice to 
smaller values with increasing $\lambda_K$. 
We have incorporated a small energy broadening to account for the finite instrumental energy resolution.}
\end{figure*}

The AFM Heisenberg-Kitaev model on the fcc lattice leads to A-type AFM, with spins in the FM plane (A-II AFM) for $J_K\!>\! 0$, consistent with the discussed magnetic order of 
La$_2B$IrO$_6$. This order persists in the regime $J_K/J_1 \to \infty$, so we focus here on the pure Kitaev interaction  (i.e., setting $J_1=J_2=0$), in order to study the effect
of quantum fluctuations around the A-II state \cite{footnote}.
Considering FM $xy$-planes stacked antiferromagnetically along $\hat{z}$, and spins making an angle $\phi$ with the $\hat{x}$-axis (Ir-O bond direction), linear spin-wave
theory leads to the dispersion
\be
\omega_\phi(\bk) \!\!=\!\! 2 J_K\! \left[(1\!+\!C^{xy}_\bk)(1\!+\!C^{xz}_\bk \! \cos^2\!\phi \!+\! C^{yz}_\bk \! \sin^2\!\phi)\right]^{1/2}
\ee
with $C^{ij}_\bk\! =\! \cos k_i \cos k_j$ ($i=x,y,z$). The zero-point energy of quantum fluctuations is $E_{\rm zp}(\phi) \!\!=\!\! \frac{1}{2} \int_\bk \omega_\phi(\bk)$ per spin
exhibits discrete minima at $\phi=n \pi/2$ ($n=0,1,2,3$). Quantum fluctuations in the presence of SOC thus break the accidental degeneracy in $\phi$
of the classical Kitaev model, favoring spins to point along the Ir-O bond directions in the FM plane. The magnon dispersion $\omega_\phi(\bk)$ in linear spin wave theory is gapless due to this classical degeneracy.

Since quantum fluctuations lift the classical XY degeneracy of the Kitaev model, we expect the concomitant development of a magnon gap in the A-II state;
indeed, magnon interaction effects, discussed below, gap out this `pseudo-Goldstone' mode.
Such an order-by-disorder gap has been discussed within different models 
for LaTiO$_3$ \cite{01_khaliullin} and the rare-earth pyrochlore Er$_2$Ti$_2$O$_7$ \cite{12_savary}.
In LaTiO$_3$, the SOC is weak, leading to a tiny gap for highly-dispersive magnons. In Er$_2$Ti$_2$O$_7$, the
order-by-disorder
gap was recently determined in INS to be $\approx 0.053$~meV  \cite{14_ross}, about an order of magnitude smaller than the
observed energy scale ($\sim\! 0.5$~meV) of its zone boundary magnetic excitations \cite{12_savary,14_ross}.

To see the emergence of this spin gap in our Kitaev model in a transparent manner, we expand the above expression for the zero point energy,
$E_{\rm zp}(\phi) \!\approx\! E_{\rm zp}(\phi=0) \!+\! \frac{1}{2} \gamma \phi^2$, where:
\be
\gamma = J_K \int_\bk  \sqrt{\frac{1+C^{xy}_\bk}{1+C^{xz}_\bk}} (C^{yz}_\bk-C^{xz}_\bk).
\ee
This leads to a pinning field $2\gamma$ in the ordered state, which is responsible for gapped magnons.
This magnon gap appears naturally within higher order spin-wave theory - incorporating magnon interactions using a self-consistent mean field theory
(see Appendix) yields a magnon gap $\approx 0.4 J_K$.  

Using this mean field approach, we have also computed the renormalized staggered 
magnetization, and find $m_{\rm AF} \approx 0.46$ in the ideal fcc lattice Kitaev model, leading to $\sim 8\%$ suppression of the classical 
$j_{\rm eff}=1/2$ order parameter due to quantum fluctuations. Taking into account the staggered octahedral rotation $\approx 11^\circ$  
in La$_2$ZnIrO$_6$ this $0.92\mu_B$ staggered magnetization translates into a uniform magnetization $\approx 0.18 \mu_B$, which 
roughly agrees with the measured value $\approx 0.22\mu_B$ in La$_2$ZnIrO$_6$ \cite{13_cao}. We have previously argued \cite{15_cook}
that the absence of a uniform magnetization in La$_2$MgIrO$_6$ might stem from the axes of staggered octahedral rotations and the stacking
direction of FM planes of the A-type AFM order being perpendicular to each other (see Fig.~5 of Ref.\cite{15_cook}); this remains to be verified
in future experiments.

Using this mean field approach, we have computed the powder averaged dynamic structure factor $S(Q,\omega)$, which is 
plotted in Fig.~\ref{theory}(f).
Our results show that the order-by-disorder gap is about a factor of $5$ smaller than the typical energy scale of
zone boundary magnon excitations. This relative scale for the gap appears to be in line with the results found using different methods for
Er$_2$Ti$_2$O$_7$ \cite{12_savary}, given the four-fold XY anisotropy for La$_2B$IrO$_6$
as opposed to the six-fold anisotropy in Er$_2$Ti$_2$O$_7$.

Since La$_2B$IrO$_6$ exhibit weak monoclinic distortions, magnon gaps may also be generated from inequivalent Kitaev couplings in 
different planes ($yz,zx,xy$). To understand the impact of such anisotropic Kitaev couplings, we have studied a set of Kitaev models with
\bea
\!\!\!\!\!\! H_{K} \!\!&=& \!\!
 J_K^x
\!\!\!  \sum_{\la \br\br'\ra_{yz}} \!\!\!\!  S^x_\br S^x_{\br'}
+ J_K \!\!\!  \sum_{\la \br\br'\ra_{xz}} \!\!\!\!  S^y_\br S^y_{\br'} +
 J_K \!\!\! \sum_{\la \br\br'\ra_{xy}}\!\!\!\! S^z_\br S^z_{\br'}
\eea
where $J_K^x = J_K (1+\lambda_K)$. Choosing $\lambda_K > 0$ then leads to moments ordered along $S_x$. 
In this case, a spin gap is already present in linear spin wave theory; incorporating magnon interactions leads to a 
slight enhancement of the spin gap. We plot the corresponding powder-averaged dynamic structure factors in Fig.~\ref{theory} for 
various values of $\lambda_K$. 
We have also computed the frustration parameter $f_K$ for this series of Heisenberg-Kitaev models in the regime where $J_K \! \gg \! J_1, J_2$, using the powder-averaged 
value of the Curie-Weiss temperature $\theta_{CW} \!=\! -J_K (1\!+\! \lambda_K/3) \!-\! 3 J_1 \!-\! 3 J_2/2$, and 
$T_N$ obtained from classical Monte Carlo simulations with scaling by $S(S\!+\!1)$ to account for quantum effects
for $S\!=\! 1/2$. 
As shown in Fig.~\ref{theory}, $f_K$ is in the range of experimental values $f \lesssim 2$.

\section{VI. Comparison with experimental data}

We next turn to a comparison between the theoretical models and the INS and thermodynamic data in order to obtain some estimates of exchange parameters.
For the $S(Q,\omega)$ plots, we use the kinematic cutoff
arising from an incident neutron energy $E_i\!=\! 7.5$~meV, which is applicable to the particular dataset that we are modeling,
and include the Ir$^{4+}$ form factor, which leads to a decay of intensity with increasing $Q$.

\begin{figure}[t]
\centering
\scalebox{0.36}{\includegraphics{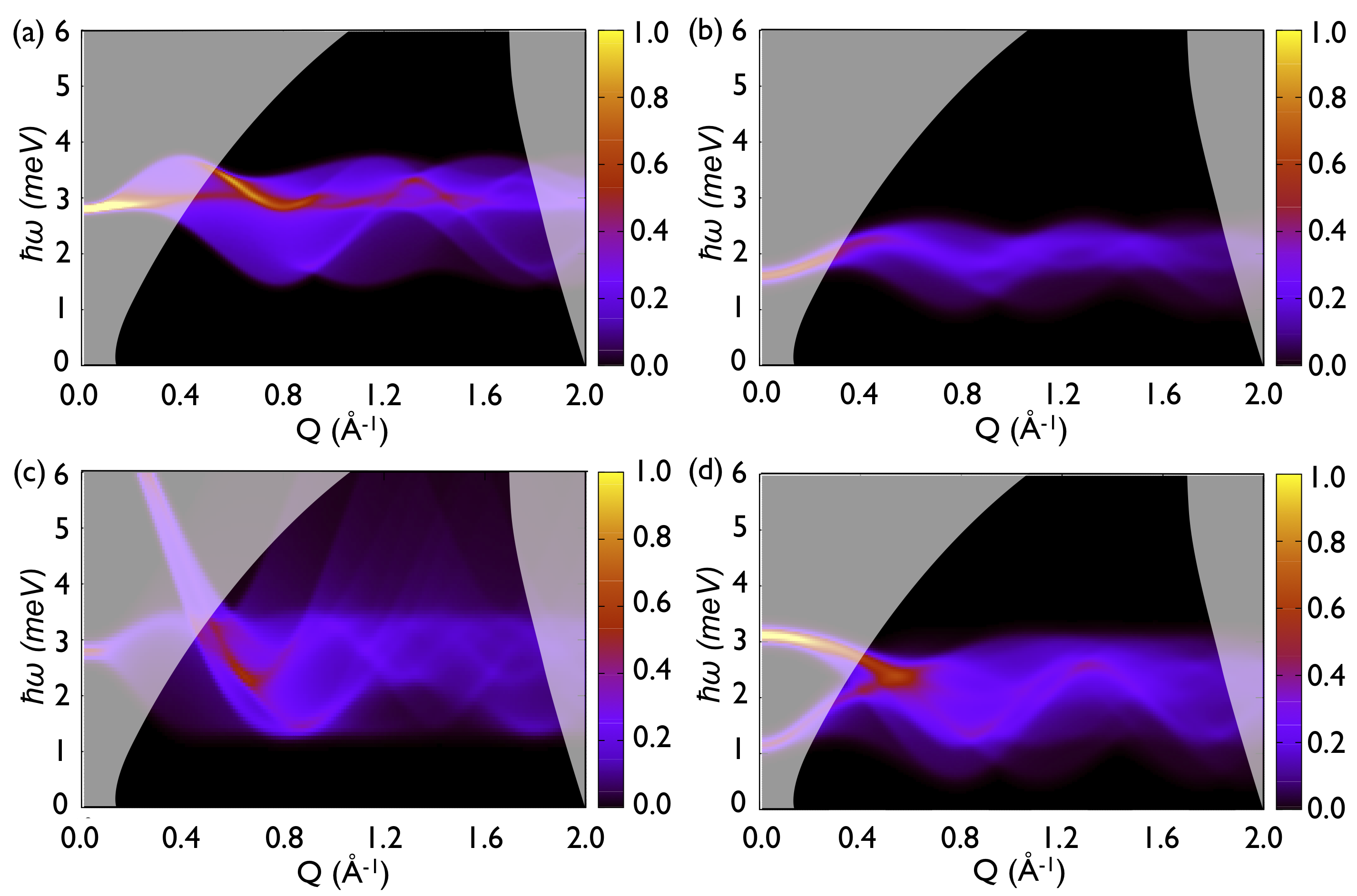}}
\caption{\label{bestsqw} (Color online) Illustrative plots of $S(Q,\omega)$ for the Heisenberg-Ising models and 
Kitaev-Heisenberg models. We have incorporated the Ir$^{4+}$ form factor, and also used the instrumental energy resolution
and the kinematic cutoff corresponding to the HYSPEC instrument with a neutron incident energy $E_i=7.5$~meV.
(a) Heisenberg-Ising model for La$_2$MgIrO$_6$ with
$J_1\!=\! 0.5$~meV, a second neighbor FM Heisenberg coupling
$J_2 \!=\! -0.3 J_1$, and uniaxial anisotropies $\lambda_1\!=\! \lambda_2\!=\! 1$. (b) Heisenberg-Ising model for
La$_2$ZnIrO$_6$, with $J_1\!=\! 0.2$~meV, $J_2\! =\! - J_1$, and $\lambda_1\!=\! \lambda_2\!=\! 1$.
(c) Kitaev-dominant model for La$_2$MgIrO$_6$ with $J_K=1.7$~meV, anisotropy $\lambda_K=0.2 J_K$, and 
$J_2=0.2 J_K$. (d) Kitaev-dominant model for La$_2$ZnIrO$_6$ with $J_K=0.7$~meV, anisotropy $\lambda_K=0.2 J_K$, and second neighbor
FM Heisenberg exchange $J_2= - 0.2 J_K$.}
\end{figure}

{\bf Heisenberg-Ising model:} We find that for  La$_2$MgIrO$_6$, choosing $J_1\!=\! 0.5$~meV, a second neighbor FM Heisenberg coupling
$J_2 \!=\! -0.3 J_1$, and uniaxial anisotropies $\lambda_1\!=\! \lambda_2\!=\! 1$, leads to a reasonable description of the INS data as shown in Fig.~\ref{bestsqw}(a).
For this choice of parameters, we obtain $\Theta_{CW} \! \approx\!  -24$K and
$T_N \! \approx \! 11$K, in good agreement with the thermodynamic data.
 For La$_2$ZnIrO$_6$, since it has a much smaller frustration parameter, we find that we have to incorporate
much larger ferromagnetic $J_2$. Choosing $J_1\!=\! 0.2$~meV , $J_2\! =\! - J_1$, and $\lambda_1\!=\! \lambda_2\!=\! 1$,
leads to $S(Q,\omega)$ shown in Fig.~\ref{bestsqw}(b). With these parameters, we obtain
$\Theta_{CW} \!\approx\! -7$K and $T_N \!\approx\! 9$K, again in good agreement with thermodynamic measurements.

{\bf Kitaev-dominant model:} For La$_2$MgIrO$_6$, choosing
$J_K\!=\! 1.7$~meV, a weak anisotropy $\lambda_K \!=\! 0.2$, and a small AFM second-neighbor Heisenberg coupling $J_2\!=\! 0.2 J_K$,
leads to reasonable agreement with the INS data as seen from Fig.~\ref{bestsqw}(c).
Using these parameters, we find $\Theta_{CW} \approx -27$K and $T_N \!\approx \! 14K$; both these results
are in good agreement with the experimental data.
We find that
incorporating a nonzero $J_1$ worsens the agreement with the data; this may possibly suggest that $J_1$ gets suppressed due to
the interfering Ir-O-O-Ir pathways.
For La$_2$ZnIrO$_6$,
we suggest that a weak ferromagnetic second neighbor exchange might be a plausible route to the smaller frustration parameter.
As shown in Fig.~\ref{bestsqw}(b), we
can get a reasonable description of the magnon dispersion with $J_K \approx 0.7$meV, $\lambda_K \!=\! 0.2$, and $J_2 = -0.2 J_K$.
In this case, we find $T_N=8$K and $\Theta_{CW}=-6K$, again in good agreement with experiments.

Interestingly, the sign change of $J_2$ between the two materials in such Kitaev-dominant models seems to correlate with expectations for DPs with magnetic 3d transition 
metals on the $B'$ site only. In these materials with lighter magnetic ions, SOC is negligible, so the Hamiltonian is generally assumed to consist of $J_1$ and $J_2$ 
Heisenberg terms only. The choice of magnetic ground state then arises from the signs and relative magnitudes of these two terms, as established previously by mean field 
theory \cite{01_lefmann}. Based on the known magnetic ground states of several materials in this family \cite{04_bos,07_holman,67_cox,14_vasala,16_koga,07_retuerto, 08_chakraborty,08_mandal,03_viola,06_martin}, we have made the empirical observation that a $d^0$ configuration for the non-magnetic ion on the B-site typically favors 
$J_2 > 0$ which agrees with the La$_2$MgIrO$_6$ result, while a $d^{10}$ configuration of the B-site favors $J_2 < 0$, as for La$_2$ZnIrO$_6$. 
We note that the Heisenberg-Ising model 
 description of the combined INS and thermodynamic data for La$_2B$IrO$_6$, which, as discussed above, requires $J_2 < 0$ for both the Mg and Zn materials, would 
not be consistent with this empirical correlation.

\section{VII. Disorder effects}


Our theoretical modeling above has considered the ideal fcc lattice or weak anisotropies induced by small non-cubic distortions. However, it is well known that $B/B'$ site mixing may play an important role in DPs \cite{06_lin, 10_aharen}. The density of antisite defects $n_d$ is commonly estimated from structural refinements of DP diffraction data, although it was previously assumed that this effect is negligible for La$_2B$IrO$_6$ in Ref.~\cite{13_cao}. We have therefore revisited the x-ray diffraction data presented in that paper and performed new structural refinements, with the site mixing included as a fitting parameter. We estimate a $B/B'$ site mixing value of $\le$~8\% and $\le$~5\% for the Mg and Zn systems respectively, with lattice constants and atomic fractional coordinates essentially identical to the values reported in Ref.~\cite{13_cao}.  Magnons scattering 
off such defects, with a magnon mean free path $\ell \sim n_d^{-1/3}$, will lead to momentum broadening $\Delta Q \sim 1/\ell$. Such disorder broadening may 
smear some of the sharp features in the theoretical $S(Q, \omega)$. Such disorder scattering, as well as
the momentum broadening arising from finite instrument resolution, also not considered in the modeling, may further improve the agreement
between theory and experiment.

\section{VIII. Conclusions}

In conclusion, we have carried out a joint experimental and theoretical investigation of the spin dynamics in the $j_{\rm eff}=1/2$ Mott
insulators La$_2B$IrO$_6$. We have shown that we can explain the combined neutron and thermodynamic data with 
conventional Heisenberg-Ising models using
significant ferromagnetic $J_2$ and a strong uniaxial Ising exchange interaction. The applicability of such models to the DP iridates is however
not clear given their small monoclinic distortions. As an alternative, we have considered models with dominant Kitaev interactions -
this unconventional form of exchange anisotropy is symmetry-allowed even on the ideal fcc lattice, and it
leads to A-type AFM order, rather than an exotic quantum spin liquid. 
Indeed, we have shown that models with dominant Kitaev interactions, supplemented by weak anisotropies and small
second neighbor couplings, appear to provide a natural alternative explanation of the INS data and thermodynamic measurements.
Experiments probing spin dynamics  in the small $Q$ regime would be valuable in distinguishing these proposals.

Going beyond these specific materials, our work points to the possibility that such Kitaev-type directional exchange couplings induced by SOC 
may be the driving force responsible for the gapped 
A-type AFM states found in a variety of other $4d/5d$-based DPs \cite{15_kermarrec, 15_taylor_1, 15_taylor_2, 89_battle, 13_carlo}. This
calls for {\it ab initio} studies of exchange interactions in such Mott insulating DP materials.
In future work, it would be useful to synthesize single crystals of
La$_2B$IrO$_6$ and other DPs. Studying the magnetic susceptibility and magnetic excitations from such single crystals might further serve to
 distinguish models with dominant Kitaev exchange
from competing conventional models with dominant uniaxial Ising interactions. Finally, our work supports the idea that materials with multiple interfering
Ir-O-O-Ir superexchange paths might be promising candidates for exploring Kitaev interactions on various lattice geometries.

\acknowledgments
We thank M.D. Lumsden, S.E. Nagler, and K.W. Plumb for useful discussions and V. O. Garlea for technical support. This research was supported by the US Department of Energy (DOE), Office of Basic Energy Sciences. Inelastic neutron scattering experiments were performed at the Spallation Neutron Source and the High Flux Isotope Reactor, which are sponsored by the Scientific User Facilities Division. A.A.A. and S.C. were supported by the Scientific User Facilities Division. G.-X.C. and D.M. were supported by the Materials Science and Engineering Division. T.J.W. acknowledges support from the Wigner Fellowship program at ORNL. A.M.C., Y.B.K., and A.P. were funded by NSERC of Canada.

\appendix

\section{\label{sec:level1} Appendix: Mean field theory of gapped magnons}
Beyond linear spin waves, we set for spin-$1/2$,
\bea
S^x_\br &=& (-1)^z (\frac{1}{2}-a^\dg_\br a^\pdg_\br) \\
S^y_\br &=& \frac{1}{2} (a^\pdg_\br + a^\dg_\br)  - \frac{1}{4} (a^\dg_\br a^\pdg_\br a^\pdg_\br + a^\dg_\br a^\dg_\br a^\pdg_\br)\\
S^z_\br &=& (-1)^z \left[ \frac{1}{2i} (a^\pdg_\br - a^\dg_\br) -  \frac{1}{4 i} (a^\dg_\br a^\pdg_\br a^\pdg_\br - a^\dg_\br a^\dg_\br a^\pdg_\br) \right]
\eea
and expand the Kitaev model Hamiltonian, only keeping terms to quartic order, which we decouple using mean field parameters $F_\bk = \la a^\dg_\bk a^\dg_{-\bk}\ra$ and $G_\bk = \la a^\dg_\bk a^\pdg_\bk \ra$. This leads to the Hamiltonian:
\bea
H_{\rm mft} = J_K \sum_{\bk>0} \begin{pmatrix} a^\dg_\bk & a^\pdg_{-\bk} \end{pmatrix} \begin{pmatrix} A_\bk & B_\bk \\ B_\bk & A_\bk \end{pmatrix}
\begin{pmatrix} a^\pdg_\bk \\ a^\dg_{-\bk} \end{pmatrix}
\eea
with $A_\bk \!=\! (2\!+\! C_{xy}\!+\!C_{xz}) \!+\! \delta \! A_\bk $,  $B_\bk \!=\! (C_{xz}\!-\!C_{xy})\! +\! \delta\! B_\bk$,
and
\bea
\delta\!A_\bk &=& 2 (\brf_{xy}-\brf_{xz})+\brf (C_{xy}-C_{xz}) - 2 \brg (C_{xy}+C_{xz}) \nonumber \\
&-& 2 (\brg_{xy}+\brg_{xz}) -4 \brg - 4 \brg_{yz} C_{yz} \\
\delta\!B_\bk &=& (\brg_{xy}-\brg_{xz}) + 2 \brg (C_{xy}-C_{xz})-\brf (C_{xy}+C_{xz}) \nonumber \\
&-& (\brf_{xy}+\brf_{xz}) - 4 \brf_{yz} C_{yz}.
\eea
Here, we have defined averages $\brf \equiv \int_\bk F_\bk$, $\brf_{ij} \equiv \int_\bk C^{ij}_\bk F_\bk$, and similarly for $G_\bk$. Requiring self-consistency, we set $F_\bk = \sinh 2\varphi_\bk$ and $G_\bk = (\cosh2\varphi_\bk-1)/2$, with the renormalized dispersion $\Omega_\bk= J_K \sqrt{A^2_\bk-B^2_\bk}$, $\cosh2\varphi_\bk = A_\bk/\Omega_\bk$, and $\sinh2\varphi_\bk = - B_\bk/\Omega_\bk$.
To solve these equations, we begin with a guess for the Hamiltonian matrix of the form $\delta\! A_\bk \!=\!  \gamma$, and $\delta \! B_\bk \!=\! 0$, where $\gamma$ represents the effect of the pinning field arising from order-by-disorder as described in the text. We then iterate the mean field equations to achieve self-consistency.

We can use these converged results to also compute the resulting dynamic structure factor, which has components
\bea
\!\!\!\!\! {\cal S}_{xx}(\bq,\omega) \!\!\! &=& \!\! \frac{1}{4} \!\int\!\! \frac{d^3\bp}{(2\pi)^3} 
(\sinh 2\varphi_\bp \sinh 2\varphi_{\bq+\bp+\bG} \nonumber \\
\!\!\!\!\!\!  &\!\!\!\!+&\!\!\!\!\! 4\! \cosh^2\!\varphi_\bp \sinh^2\!\varphi_{\bq+\bp+\bG})
\delta(\omega \! -\! \Omega_\bp \!-\! \Omega_{\bq+\bp+\bG}) 
\nonumber \\
\!\!\!\!\!\! {\cal S}_{yy}(\bq,\omega) \!\!\! &=&\!\! \! (\cosh\!2\varphi_\bq\!+\! \sinh\!2\varphi_\bq) (1\!-\! 2\brg\! -\! \brf) \delta(\omega\!-\!\Omega_\bq) \nonumber \\
\!\!\!\!\!\! {\cal S}_{zz}(\bq,\omega) \!\!\! &=&\!\! \! (\cosh\!2\varphi_\bq\!-\! \sinh\!2\varphi_\bq) (1\!-\! 2\brg\! + \! \brf) \delta(\omega\!-\!\Omega_{\bq\!+\!\bG}) \nonumber
\eea
where $\bG \equiv (0,0,\pi)$. The first term corresponds to longitudinal fluctuations while the latter two correspond to transverse fluctuations. We find, numerically, that the longitudinal fluctuations make a very small contribution to the structure factor, and can be ignored in practice. Powder averaging leads to $S(Q,\omega)$, with $Q=|\vec{q}|$, which we convolute with a Gaussian function representing the instrumental energy resolution, and plot in Fig.~\ref{theory}(f) above.


\begin{thebibliography}{99}
\bibitem{note}Ref.~\cite{15_zhu} suggests that there are actually two different magnetic transition temperatures in close proximity for La$_2$ZnIrO$_6$ ($T_{1}$~$=$~7.3~K and $T_{2}$~$=$~8.5~K), but this finding is not important for the results presented in our work.
\bibitem{footnote} While a small $J_1\!>\! 0$ appears to be ultimately necessary for globally favoring the AFM-II state in the classical Heisenberg-Kitaev model, 
 it does not qualitatively or quantitatively affect the spin-wave energetics
 around the A-II state for $J_1 \ll J_K$.
\end{thebibliography}
\end{document}